\documentclass[seceq,preprint]{ptptex}
\usepackage{wrapft}
\usepackage{graphicx}
\newcommand{\ket}[1]{| {#1} \rangle}     %%
\newcommand{\bbra}[1]{\langle\!\langle {#1} |}     %%
\newcommand{\kket}[1]{| {#1} \rangle\!\rangle}     %%
\newcommand{\rkket}[1]{| {#1} )\!)}     %%
\newcommand{\rdket}[1]{|\!| {#1} )}     %%
\newcommand{\dket}[1]{|\!| {#1} \rangle}     %%
\newcommand{\wtilde}[1]{\widetilde{#1}} %%

\def\<{\langle}
\def\>{\rangle}
\def\bsub{\begin{subequations}}
\def\esub{\end{subequations}}
\def\beqn{\begin{eqnarray}}
\def\eeqn{\end{eqnarray}}
\def\b{\begin{equation}}

\title{%        %You can use \\ for explicit line-break
Boson Realization of the $su(3)$-Algebra. III
}
\subtitle{
Schwinger Representation for the Elliott Model
}    %use this when you want a subtitle

\author{%       %Use \sc for the family name
Constan\c{c}a {\sc Provid\^encia},$^{1}$
Jo\~ao da {\sc Provid\^encia},$^{1}$\\
Yasuhiko {\sc Tsue}$^{2}$ 
and Masatoshi {\sc Yamamura}$^{3}$
}

\inst{%         %Affiliation, neglected when [addenda] or [errata]
$^{1}$Departamento de Fisica, Universidade de Coimbra, 3004-516 Coimbra, 
Portugal\\
$^2$Physics Division, Faculty of Science, Kochi University, Kochi 780-8520, 
Japan\\
$^{3}$Faculty of Engineering, Kansai University, Suita 564-8680, Japan
}

\recdate{%      %Editorial Office will fill in this.
\today
}

\abst{%       %this abstract is neglected when [addenda] or [errata]
Following the basic idea developed in (I), the $su(3)$-model 
presented by Elliott is investigated. 
A method for constructing the orthogonal set, in which the angular momenta 
are good quantum numbers, is discussed without using the angular 
momentum projection.
}
\begin{document}
\maketitle

\section{Introduction}

In (I) and (II), we presented two forms for the boson representation of the 
$su(3)$-algebra.\cite{1} 
They are extensions from the Schwinger\cite{2} 
and the Holstein-Primakoff\cite{3} representation of the $su(2)$-algebra 
and both may be useful for the study of boson realization of the Lipkin 
model for many-fermion system. 
Of course, the original Lipkin model is based on the $su(2)$-algebra.\cite{4}
For the completeness of the papers related to the $su(3)$-algebra, 
we must contact the $su(3)$-model presented by Elliott in 1958.\cite{5}

As is well known, the $su(3)$-model presented by Elliott brought in a 
new phase in nuclear theory, especially, in the study of nuclear structure 
theory. 
It gives us the Lie algebraic approach to the study of collective motion. 
We can find some historical examples of the studies based on the Lie algebra 
in Ref.\citen{6}. 
Much more detailed references until the begining of 1990s are 
listed in Ref.\citen{7} and more recent information can be found 
in Ref.\citen{8}. 
The Elliott $su(3)$-model, of course, consists of eight generators, in which 
the five and the remaining three are related to the quadrupole moment and the 
orbital angular momentum, respectively. 
A characteristic point of this model is as follows: 
If the Casimir operator of the $su(3)$-algebra is related to the 
Hamiltonian, we can easily derive the rotational spectra and this Hamiltonian 
becomes the origin of the quadrupole-quadrupole interaction. 
However, it is very tedious to obtain the eigenstates of the Hamiltonian. 
Of course, by making an appropriate linear combination for the 
original eight generators, we obtain the form presented in (I) and 
following the idea in (I), the eigenstates are obtained. 
But, this orthogonal set is not expressed in terms of the eigenstates for 
the orbital angular momentum. 
Then, by the projection method, we derive the final form. 
This procedure was presented by Elliott and, in Appendix of this paper, 
we show the form.

A main aim of this paper is to demonstrate an idea, with the help of 
which we can arrive at the eigenstates with respect to the 
angular momentum without making a detour such as angular momentum projection. 
The basic idea is borrowed from that presented in (I): 
The excited state generating operators play a role of connecting the states 
with the minimum weight for the angular momentum. 
This has been in \S 5 of (I). 
Contrary to the form treated in (I), we cannot find six hermitian operators 
to specify the orthogonal set. 
Therefore, a special device is necessary.

In the next section, the starting framework of the Elliott $su(3)$-model is 
given. 
Section 3 is devoted to constructing the intrinsic state. 
In \S 4, the structure of the excited states is discussed. 
In \S 5, various properties are mentioned, especially, an idea constructing 
the excited states is introduced. 
In Appendix, the angular momentum projection method is treated.

\section{The $su(3)$-algebra in the form suitable for the Elliott model 
and its Schwinger representation}

The $su(3)$-model proposed by Elliott is composed of the orbital angular 
momentum and the modified quadrupole moment operators which are denoted 
as ${\hat L}_{\pm,0}$ and ${\hat Q}_{\pm 2, \pm 1,0}$, respectively. 
These operators can be expressed in terms of
\break 
$({\hat S}^1,{\hat S}_1,{\hat S}^2,{\hat S}_2,{\hat S}_2^1,{\hat S}_1^2,
{\hat S}_1^1,{\hat S}_2^2)$: 
\bsub\label{2-1}
\beqn
& &{\hat L}_+=\sqrt{2}({\hat S}_1+{\hat S}^2) \ , \qquad
{\hat L}_-=\sqrt{2}({\hat S}^1+{\hat S}_2) \ , \qquad
{\hat L}_0={\hat S}_2^2-{\hat S}_1^1 \ , 
\label{2-1a}\\
& &{\hat Q}_0={\hat S}_2^2+{\hat S}_1^1 \ , 
\qquad\label{2-1b}\\
& &{\hat Q}_1=\sqrt{3}({\hat S}_1-{\hat S}^2) \ , \qquad
{\hat Q}_2=\sqrt{6}{\hat S}_2^1 \ , \nonumber\\
& &{\hat Q}_{-1}=-\sqrt{3}({\hat S}^1-{\hat S}_2) \ , \qquad
{\hat Q}_{-2}=\sqrt{6}{\hat S}_1^2 \ . 
\label{2-1c}
\eeqn
In order to descriminate the role of ${\hat Q}_{\pm 1}$ and ${\hat Q}_{\pm 2}$ 
from that of ${\hat Q}_0$ explicitely, we use the following notations: 
\begin{equation}\label{2-1d}
{\hat C}_1^*={\hat Q}_1 \ , \qquad 
{\hat C}_2^*={\hat Q}_2 \ , \qquad 
{\hat C}_1=-{\hat Q}_{-1} \ , \qquad 
{\hat C}_2={\hat Q}_{-2} \ .
\end{equation}
\esub
Formally, ${\hat L}_{\pm,0}$, ${\hat Q}_0$, ${\hat C}_{1,2}^*$ and 
${\hat C}_{1,2}$ correspond to ${\hat I}_{\pm,0}$, ${\hat M}_0$, 
${\hat D}_{\pm}^*$ and ${\hat D}_{\pm}$, respectively. 
However, the algebraic structures of these operators are different of each 
other. 
This can be seen in the following commutation relations: 
\bsub\label{2-2}
\beqn
& &[\ {\hat L}_+\ , \ {\hat L}_- \ ]=2{\hat L}_0 \ , \qquad 
[\ {\hat L}_0\ , \ {\hat L}_{\pm} \ ]=\pm{\hat L}_\pm \ , 
\label{2-2-a}\\
& &[\ {\hat L}_+\ , \ {\hat Q}_0 \ ]=\sqrt{6}{\hat C}_1^* \ , \qquad
[\ {\hat L}_-\ , \ {\hat Q}_0 \ ]=-\sqrt{6}{\hat C}_1 \ , \qquad
[\ {\hat L}_0\ , \ {\hat Q}_0 \ ]=0 \ , 
\label{2-2b}\\
& &[\ {\hat L}_+\ , \ {\hat C}_2^* \ ]=0 \ , \quad
[\ {\hat L}_+\ , \ {\hat C}_1^* \ ]=2{\hat C}_2^* \ , \quad
[\ {\hat L}_-\ , \ {\hat C}_2^* \ ]=2{\hat C}_1^* \ , \quad
[\ {\hat L}_-\ , \ {\hat C}_1^* \ ]=\sqrt{6}{\hat Q}_0 \ , \nonumber\\
& &[\ {\hat L}_0\ , \ {\hat C}_2^* \ ]=2{\hat C}_2^* \ , \qquad
[\ {\hat L}_0\ , \ {\hat C}_1^* \ ]={\hat C}_1^* \ , 
\label{2-2c}\\
& &[\ {\hat Q}_0\ , \ {\hat C}_2^* \ ]=0 \ ,
[\ {\hat Q}_0\ , \ {\hat C}_1^* \ ]=-3\sqrt{3/2}{\hat L}_+ \ , \nonumber\\
& &[\ {\hat C}_2^*\ , \ {\hat C}_1^* \ ]=0 \ , \qquad
[\ {\hat C}_2^*\ , \ {\hat C}_1 \ ]=3{\hat L}_+ \ ,  
\label{2-2d}\\
& &[\ {\hat C}_2^*\ , \ {\hat C}_2 \ ]=6{\hat L}_0 \ , \qquad 
[\ {\hat C}_1^*\ , \ {\hat C}_1 \ ]=3{\hat L}_0 \ . 
\label{2-2e} 
\eeqn
\esub
We can see in the relation that the set $({\hat L}_{\pm,0})$ forms the 
$so(3)$-algebra, but, contrary to ${\hat M}_0$ and $({\hat D}_\pm)$ in (I), 
${\hat Q}_0$ and $({\hat C}_1^*,{\hat C}_2^*)$ do not form spherical 
tensors with respect to $({\hat L}_{\pm.0})$. 

The Casimir operator ${\hat \Gamma}_{su(3)}$ is expressed in the form 
\beqn\label{2-3}
{\hat \Gamma}_{su(3)}&=&
(1/6)({\hat Q}_2{\hat Q}_{-2}-{\hat Q}_1{\hat Q}_{-1}+{\hat Q}_0^2
-{\hat Q}_{-1}{\hat Q}_{1}+{\hat Q}_{-2}{\hat Q}_2) \nonumber\\
& &+(1/2)[{\hat L}_0^2+(1/2)({\hat L}_+{\hat L}_-+{\hat L}_-{\hat L}_+)] \ .
\eeqn
The form (\ref{2-3}) can be rewritten as 
\beqn
& &{\hat \Gamma}_{su(3)}
=(1/2)\left[
(1/3)
({\hat {\mib C}}^*{\hat {\mib C}}+{\hat {\mib C}}{\hat {\mib C}}^*)
+(1/3){\hat Q}_0^2+{\hat {\mib L}}^2\right] \ , 
\label{2-4}\\
& &{\hat {\mib C}}^*{\hat {\mib C}}
={\hat C}_2^*{\hat C}_2+{\hat C}_1^*{\hat C}_1 \ , \qquad
{\hat {\mib C}}{\hat {\mib C}}^*
={\hat C}_2{\hat C}_2^*+{\hat C}_1{\hat C}_1^* \ , 
\nonumber\\
& &{\hat {\mib L}}^2={\hat L}_0^2+(1/2)
({\hat L}_+{\hat L}_-+{\hat L}_-{\hat L}_+) \ . 
\label{2-5}
\eeqn

Associating the above $su(3)$-algebra, the $su(1,1)$-algebra also plays a 
central role in this paper. 
It is composed of three generators which were denoted as 
$({\hat T}^1,{\hat T}_1,{\hat T}_1^1)$ in 
Ref.\citen{A}, hereafter, referred to as (A). 
In this paper, we use the notation $({\wtilde T}_{\pm,0})$ which is denoted as 
\begin{equation}\label{2-6}
{\wtilde T}_+={\hat T}^1 \ , \qquad 
{\wtilde T}_-={\hat T}_1 \ , \qquad
{\wtilde T}_0=(1/2){\hat T}_1^1 \ . 
\end{equation}
The commutation relation and the Casimir operator are given as follows: 
\beqn
& &[\ {\wtilde T}_+\ , \ {\wtilde T}_- \ ]=-2{\wtilde T}_0 \ , \qquad 
[\ {\wtilde T}_0\ , \ {\wtilde T}_{\pm} \ ]=\pm{\wtilde T}_\pm \ , 
\label{2-7}\\
& &{\hat \Gamma}_{su(1,1)}=2{\wtilde {\mib T}}^2
=2\left[{\wtilde T}_0^2-(1/2)
({\wtilde T}_+{\wtilde T}_- +{\wtilde T}_-{\wtilde T}_+)\right] \ . 
\label{2-8}
\eeqn

A possible boson realization of the above two algebras were obtained 
in terms of six kinds of boson operators $({\hat a}_{\pm,0},{\hat b}_{\pm,0})$ 
and $({\hat a}_{\pm,0}^*,{\hat b}_{\pm,0}^*)$. 
The detail can be found in (A). 
By changing the notations adopted in (A) such as 
${\hat a}^1 \rightarrow {\hat b}_0$, ${\hat a}_2 \rightarrow {\hat a}_+$, 
${\hat a}_1 \rightarrow {\hat a}_-$, ${\hat b} \rightarrow {\hat a}_0$, 
${\hat b}_1^1 \rightarrow {\hat b}_+$, ${\hat b}_2^1 \rightarrow {\hat b}_-$, 
we have the following form: 
\bsub\label{2-9}
\beqn
& &{\hat L}_+=\sqrt{2}[({\hat a}_+^*{\hat a}_0 + {\hat a}_0^*{\hat a}_-)
+({\hat b}_+^*{\hat b}_0 + {\hat b}_0^*{\hat b}_-)] \ , \nonumber\\
& &{\hat L}_-=\sqrt{2}[({\hat a}_0^*{\hat a}_+ + {\hat a}_-^*{\hat a}_0)
+({\hat b}_0^*{\hat b}_+ + {\hat b}_-^*{\hat b}_0)] \ , \nonumber\\
& &{\hat L}_0=({\hat a}_+^*{\hat a}_+ - {\hat a}_-^*{\hat a}_-)
+({\hat b}_+^*{\hat b}_+ - {\hat b}_-^*{\hat b}_-) \ , 
\label{2-9a}\\
& &{\hat Q}_0=[(-2{\hat a}_0^*{\hat a}_0 + 
{\hat a}_+^*{\hat a}_+ + {\hat a}_-^*{\hat a}_-)
+(2{\hat b}_0^*{\hat b}_0 - 
{\hat b}_+^*{\hat b}_+ - {\hat b}_-^*{\hat b}_-)] \ , \qquad\ 
\label{2-9b}\\
& &{\hat C}_1^*=\sqrt{3}[({\hat a}_0^*{\hat a}_- - {\hat a}_+^*{\hat a}_0)
-({\hat b}_0^*{\hat b}_- - {\hat b}_+^*{\hat b}_0)] \ , \nonumber\\
& &{\hat C}_2^*=\sqrt{6}({\hat a}_+^*{\hat a}_- - {\hat b}_+^*{\hat b}_-) \ , 
\nonumber\\
& &{\hat C}_1=\sqrt{3}[({\hat a}_-^*{\hat a}_0 - {\hat a}_0^*{\hat a}_+)
-({\hat b}_-^*{\hat b}_0 - {\hat b}_0^*{\hat b}_+)] \ , \nonumber\\
& &{\hat C}_2=\sqrt{6}({\hat a}_-^*{\hat a}_+ - {\hat b}_-^*{\hat b}_+) \ , 
\label{2-9c}
\eeqn
\esub
\vspace{-0.9cm}
\beqn\label{2-10}
& &{\wtilde T}_+={\hat a}_0^*{\hat b}_0^* - 
{\hat a}_+^*{\hat b}_-^* - {\hat a}_-^*{\hat b}_+^* \ , \nonumber\\
& &{\wtilde T}_-={\hat b}_0{\hat a}_0 - 
{\hat b}_-{\hat a}_+ - {\hat b}_+{\hat a}_- \ , \nonumber\\
& &{\wtilde T}_0=(1/2)[({\hat a}_0^*{\hat a}_0 + 
{\hat b}_0^*{\hat b}_0) + ({\hat a}_+^*{\hat a}_+ + {\hat b}_-^*{\hat b}_-) 
+ 
({\hat a}_-^*{\hat a}_- + {\hat b}_+^*{\hat b}_+)+3] \ . 
\eeqn

The commutation relations for $({\hat L}_{\pm,0})$ defined in the 
relation (\ref{2-9a}) tell us that the boson operators $({\hat a}_{\pm,0}^*)$ 
and $({\hat b}_{\pm,0}^*)$ are vectors (rank $=1$). 
Further, it may be important to see that the expressions (\ref{2-9}) 
and (\ref{2-10}) give us the relation 
\begin{equation}\label{2-11}
[\ {\rm any\ of\ the}\ su(3)\hbox{\rm -generators}\ , \ 
{\rm any\ of\ the}\ su(1,1)\hbox{\rm -generators}\ ]=0 \ . 
\end{equation}
The Casimir operator (\ref{2-4}) can be reexpressed in the form 
\beqn
& &{\hat \Gamma}_{su(3)}=
2\left[{\hat {\mib T}}^2-3/4+(1/3)({\wtilde R}_0)^2\right] \ , 
\label{2-12}\\
& &{\wtilde R}_0=(1/2)[
({\hat a}_0^*{\hat a}_0 - 
{\hat b}_0^*{\hat b}_0) + ({\hat a}_+^*{\hat a}_+ - {\hat b}_+^*{\hat b}_+) 
+ 
({\hat a}_-^*{\hat a}_- - {\hat b}_-^*{\hat b}_-)] \ . 
\label{2-13}
\eeqn
Naturally, we have 
\begin{equation}\label{2-14}
[\ {\wtilde R}_0 \ , \ 
{\rm any\ of\ the}\ su(3)\hbox{\rm -\ and the}\ su(1,1)\hbox{\rm -generators}\ ]
=0 \ . 
\end{equation}
The proof of the relation (\ref{2-12}) is straightforward, but tedious. 
The above is an outline of the $su(3)$-algebra and its associating 
$su(1,1)$-algebra in Schwinger representation, which may be suitable for 
analyzing the Elliott model.

\section{Construction of the intrinsic state}

The present system consists of six kinds of boson operators and it is easily 
found that five operators ${\wtilde {\mib T}}^2$, ${\wtilde T}_0$, 
${\wtilde R}_0$, ${\hat {\mib L}}^2$ and ${\hat L}_0$ are mutually commuted. 
However, the operator ${\hat Q}_0$ cannot play the same role as that of 
${\hat M}_0$ in (I). 
Therefore, in order to obtain the orthogonal set for the present system, 
a special device may be necessary. 
This is essentially different from (I). 
However, even for this task, the determination of the intrinsic state 
given in (A) must be performed. 

As was discussed in (A), the intrinsic state, which we denote $\ket{m}$, 
should obey the following condition: 
\begin{equation}\label{3-1}
{\hat C}_1\ket{m}={\hat C}_2\ket{m}={\hat L}_-\ket{m}=0 \ , \qquad
{\wtilde T}_-\ket{m}=0\ . 
\end{equation}
In (I), the relation corresponding to the condition (\ref{3-1}) was given as 
${\hat D}_-\ket{m}={\hat D}_+\ket{m}={\hat I}_-\ket{m}=0$ and 
${\wtilde T}_-\ket{m}=0$. 
Further, $\ket{m}$ should be an eigenstate for ${\wtilde T}_0$, 
${\wtilde R}_0$ and ${\hat L}_0$. 
The state $\ket{m}$ which satisfies the condition (\ref{3-1}) is easily 
obtained:
\beqn\label{3-2}
\ket{m}=\ket{T,R}&=&\sqrt{(T-3/2+R)!(T-3/2-R)!}
({\hat a}_-^*)^{T-3/2+R}({\hat b}\-^*)^{T-3/2-R}\ket{0}\ . \qquad
\eeqn
It satisfies the eigenvalue equations 
\bsub\label{3-3}
\beqn
& &{\wtilde T}_0\ket{T,R}=T\ket{T,R}\ , 
\label{3-3a}\\
& &{\wtilde R}_0\ket{T,R}=R\ket{T,R}\ , 
\label{3-3b}\\
& &{\hat L}_0\ket{T,R}=-L^0\ket{T,R}\ , \qquad L^0=2(T-3/2) \ . 
\label{3-3c}
\eeqn
Further, $\ket{T,R}$ is also the eigenstate of ${\hat Q}_0$: 
\begin{equation}\label{3-3d}
{\hat Q}_0\ket{T,R}=Q^0\ket{T,R}\ , \qquad 
Q^0=2R \ .
\end{equation}
\esub
In the same region as that shown in (I), $T$ runs in the region 
\begin{equation}\label{3-4}
T=3/2\ , \ 2\ , \ 5/2\ , \ 3\ , \cdots \ . \quad (T\geq 3/2)
\end{equation}
Further, $R$ runs in the region 
\begin{eqnarray}\label{3-5}
& &R=-(T-3/2)\ , \ -(T-3/2)+1\ , \cdots \ , (T-3/2)-1 \ , \ (T-3/2)\ . 
\nonumber\\
& &\qquad\qquad (-(T-3/2) \leq R \leq T-3/2)
\end{eqnarray}
The eigenvalues of ${\hat \Gamma}_{su(3)}$ and ${\hat \Gamma}_{su(1,1)}$ 
for the state $\ket{T,R}$ are expressed in the same forms as (I): 
\begin{equation}\label{3-6}
\hbox{\rm the\ eigenvalue\ of\ }
\begin{cases} {\hat \Gamma}_{su(3)}=2\left[(T-3/2)(T-3/2+2)+(1/3)R^2
\right] \ , \\
{\hat \Gamma}_{su(1,1)}=2T(T-1) \ . 
\end{cases}
\end{equation}
In (I), we showed three cases for specifying the state $\ket{m}$. 
However, in the present case, it may be enough to consider the case $(T,R)$, 
because $L^0$ and $Q^0$ are simply proportional to $(T-3/2)$ and $R$, 
respectively. 
Later, instead of $(T,R)$, we will use $(l_a,l_b)$, where $l_a$ and $l_b$ 
denote numbers of the bosons ${\hat a}_-^*$ and ${\hat b}_-^*$ in the 
state $\ket{T,R}$, respectively: 
\bsub\label{3-8}
\beqn
& &l_a=T-3/2+R\ , \qquad l_b=T-3/2-R \ , 
\label{3-8a}
\eeqn
namely, 
\beqn\label{3-8b}
& &T-3/2=(1/2)(l_a+l_b) \ , \qquad R=(1/2)(l_a-l_b) \ . 
\eeqn
\esub
The eigenvalues $L^0$ and $Q^0$ shown in the relations (\ref{3-3c}) and 
(\ref{3-3d}) are given as 
\begin{equation}\label{3-9}
L^0=l_a+l_b \ , \qquad 
Q^0=l_a-l_b \ . 
\end{equation}
Of course, $(l_a,l_b)$ plays the same role as that of $(m_0,m_1)$ in (I).

\section{Structure of the excited states}

In \S\S 4 and 5 of 
(I), we showed two forms of the orthogonal sets of the $su(3)$-algebra 
suitable for the Lipkin model. 
First is based on the fact that the excited state generating operators 
${\hat D}_\pm^*$ are regarded as spherical tensors with rank 1/2 and 
$z$-components 1/2 and $-1/2$. 
Second is not based on this fact, but of a form in which the minimum weight 
states for the $su(2)$-spin are connected through certain operators 
composing the $su(3)$-generators. 
In the present case, the excited state generating operators 
${\hat C}_1^*$ and ${\hat C}_2^*$ do not form spherical tensors for 
$({\hat L}_{\pm,0})$. 
Then, by adopting the basic idea for the second form, we will show 
the orthogonal set for the $su(3)$-algebra suitable for the Elliott model.

First, let us introduce the following operators: 
\bsub\label{4-1}
\beqn
{{\mib C}}_1^*&=&
{\hat C}_1^*{\hat L}_0({\hat L}_0-1/2)({\hat L}_0-1)
+\sqrt{3/2}{\hat L}_+{\hat Q}_0({\hat L}_0-1/2)({\hat L}_0-1)\nonumber\\
& &-(3/4){\hat L}_+^2{\hat C}_1({\hat L}_0-1)+(1/4){\hat L}_+^3{\hat C}_2 \ , 
\label{4-1a}\\
{{\mib C}}_2^*&=&
{\hat C}_2^*({\hat L}_0-1/2){\hat L}_0({\hat L}_0+1/2)({\hat L}_0+1)\nonumber\\
& &+{\hat L}_+{\hat C}_1^*({\hat L}_0-1/2){\hat L}_0({\hat L}_0+1/2)
+\sqrt{3/8}{\hat L}_+^2{\hat Q}_0({\hat L}_0-1/2){\hat L}_0\nonumber\\
& &-(1/4){\hat L}_+^3{\hat C}_1({\hat L}_0-1/2)
+(1/16){\hat L}_+^4{\hat C}_2 \ .
\label{4-1b}
\eeqn
\esub
They satisfy the relations 
\bsub\label{4-2}
\beqn
& &[\ {\hat L}_- \ , \ {{\mib C}}_1^*\ ]=
\left[(3/2){\hat C}_1^{*2}{\hat L}_0^2+
\sqrt{3/2}{\hat L}_+{\hat Q}_0(2{\hat L}_0-1/2)
-(3/4){\hat L}_+^2{\hat C}_1\right]\cdot {\hat L}_- \ , 
\nonumber\\
& &[\ {\hat L}_0 \ , \ {{\mib C}}_1^*\ ]={\mib C}_1^*\ , 
\label{4-2a}\\
& &[\ {\hat L}_- \ , \ {{\mib C}}_2^*\ ]=
\Big[{\hat C}_2^{*}({\hat L}_0+1)(2{\hat L}_0+3/2)(2{\hat L}_0+1)
+3{\hat L}_+{\hat C}_1^*({\hat L}_0+1/2)^2\nonumber\\
& &\qquad\qquad\qquad
+\sqrt{3/8}{\hat L}_+^2{\hat Q}_0(2{\hat L}_0+1/2)-(1/4){\hat L}_+^3{\hat C}_1
\Big]\cdot{\hat L}_- \ , 
\nonumber\\
& &[\ {\hat L}_0 \ , \ {{\mib C}}_2^*\ ]=2{\mib C}_2^*\ , 
\label{4-2b}
\eeqn
\esub
\vspace{-0.8cm}
\begin{equation}\label{4-3}
{\mib C}_1\ket{T,R}=0 \ , \qquad {\mib C}_2\ket{T,R}=0 \ . \qquad\qquad\qquad
\qquad\qquad\qquad\qquad\quad
\end{equation}
We, further, introduce the operator ${\mib Q}_0$ in the form 
\beqn\label{4-4}
{\mib Q}_0&=&
{\hat Q}_0({\hat L}_0-1)({\hat L}_0-3/2)-\sqrt{3/2}
({\hat L}_+{\hat C}_1+{\hat C}_1^*{\hat L}_-)({\hat L}_0-3/2)\nonumber\\
& &+\sqrt{3/8}({\hat L}_+^2{\hat C}_2+{\hat C}_2^*{\hat L}_-^2)\ . 
\eeqn
The operator ${\mib Q}_0$ obeys 
\bsub\label{4-5}
\beqn
& &[\ {\hat L}_- \ , \ {{\mib Q}}_0\ ]=
-\left[{\hat Q}_0{\hat L}_0+\sqrt{3/8}({\hat L}_+{\hat C}_1+{\hat C}_1^*
{\hat L}_-)\right]\cdot{\hat L}_- \ , 
\label{4-5a}\\
& &[\ {\hat L}_0 \ , \ {{\mib Q}}_0\ ]=0 \ . 
\label{4-5b}
\eeqn
\esub
Since ${\hat Q}_0$ and $({\hat L}_+{\hat C}_1+{\hat C}_1^*{\hat L}_-)$ 
commute with ${\hat L}_0$, ${\mib Q}_0$ is hermitian: 
\begin{equation}\label{4-6}
{\mib Q}_0^*={\mib Q}_0 \ . 
\end{equation}

Under the above preparation, we define the following state: 
\begin{equation}\label{4-7}
\kket{nR,L;T}=({\mib C}_1^*)^{2(T-3/2)-L-2n}({\mib C}_2^*)^n\ket{T,R} \ .
\end{equation}
Here, $n$ denotes positive integer obeying 
\begin{equation}\label{4-8}
n=0,\ 1,\ 2,\cdots , \ (T-3/2)-L/2 \ . \qquad 
(0\leq (T-3/2)-L/2)
\end{equation}
It is easily verified that the state (\ref{4-7}) obeys 
\bsub\label{4-9}
\beqn
& &{\wtilde T}_-\kket{nR,L;T}=0 \ , 
\label{4-9a}\\
& &{\wtilde T}_0\kket{nR,L;T}=T\kket{nR,L;T} \ , 
\label{4-9b}
\eeqn
\esub
\vspace{-0.8cm}
\begin{equation}\label{4-10}
{\wtilde R}_0\kket{nR,L;T}=R\kket{nR,L;T}\ , 
\end{equation}
\vspace{-0.8cm}
\bsub\label{4-11}
\beqn
& &\ \ {\hat L}_-\kket{nR,L;T}=0 \ , 
\label{4-11a}\\
& &\ \ {\hat L}_0\kket{nR,L;T}=-L\kket{nR,L;T} \ . 
\label{4-11b}
\eeqn
\esub
From the relations (\ref{4-9})$\sim$(\ref{4-10}), we are able to obtain the 
eigenstate for ${\wtilde {\mib T}}^2$, ${\wtilde T}_0$, ${\wtilde R}_0$, 
${\hat {\mib L}}^2$ and ${\hat L}_0$ with the eigenvalues 
$T(T-1)$, $T_0$, $R$, $L(L+1)$ and $L_0$, respectively, in the form 
\beqn\label{4-12}
& &\kket{nR,LL_0;TT_0}=({\wtilde T}_+)^{T_0-T}
({\hat L}_+)^{L+L_0}\kket{nR,L;T} \ . \nonumber\\
& &\qquad
T_0=T,\ T+1,\ T+2,\ \cdots \ , \qquad
L_0=-L,\ -L+1,\ \cdots ,\ L-1,\ L\ . \ \ 
\eeqn
However, it should be noted that $n$ is not a quantum number, i.e., for 
$n'\neq n$, \break
$\bbra{n'R,LL_0;TT_0}nR,LL_0;TT_0\rangle\!\rangle\neq 0$, 
but, the set $\{\kket{nR,LL_0;TT_0}\}$ is linearly independent. 
Therefore, by an appropriate method, for example, the Schmidt method, 
we have an orthogonal set, which we denote $\{\kket{kR,LL_0;TT_0}\}$:
\bsub\label{4-13}
\beqn
& &\bbra{k'R,LL_0;TT_0}kR,LL_0;TT_0\rangle\!\rangle=0 \ . \qquad
({\rm for}\ k'\neq k)
\label{4-13a}
\eeqn
If the relation (\ref{4-13a}) holds, we have 
\beqn
& &\bbra{k'R,L;T}kR,L;T\rangle\!\rangle=0 \ . \qquad
({\rm for}\ k'\neq k)
\label{4-13b}
\eeqn
\esub
Here, $\kket{kR,L;T}$ is defined through 
\begin{equation}\label{4-14}
\kket{kR,LL_0;TT_0}=({\hat T}_+)^{T_0-T}({\hat L}_+)^{L+L_0}
\kket{kR,L;T}\ . 
\end{equation}
The relation (\ref{4-13b}) tells us that the set $\{\kket{kR,L;T}\}$ 
is also an orthogonal set. 

We will present a possible idea for obtaining the orthogonal set 
which is related to ${\mib Q}_0$. 
We note that the operator ${\mib Q}_0$ defined in the relation (\ref{4-4}) 
commutes with ${\wtilde T}_0$, ${\wtilde R}_0$ and ${\hat L}_0$. 
The properties of ${\mib Q}_0$ shown in the relation (\ref{4-5a}) and the 
relation (\ref{4-11}) give us 
\begin{equation}\label{4-15}
{\hat L}_-{\mib Q}_0\kket{kR,L;T}=0 \ . 
\end{equation}
The relation (\ref{4-15}) tells us that ${\mib Q}_0\kket{kR,L;T}$ can be 
expressed in terms of linear combination for 
$\kket{kR,L;T}$ with different $k$: 
\begin{equation}\label{4-16}
{\mib Q}_0\kket{kR,L;T}=\sum_{k'}C_{kk'}\kket{k'R,L;T} \ . 
\end{equation}
The relation (\ref{4-16}) permits us to set up the following 
eigenvalue equation: 
\beqn
& &{\mib Q}_0\kket{QR,L;T}=Q(2T-2)(2T-3/2)\kket{QR,L;T}\ , 
\label{4-17}\\
& &\kket{QR,L;T}=\sum_{k}C_k(Q)\kket{kR,L;T}\ . 
\label{4-18}
\eeqn
After solving the eigenvalue equation (\ref{4-17}), we obtain 
\begin{equation}\label{4-19}
\kket{QR,LL_0;TT_0}=({\wtilde T}_+)^{T_0-T}({\hat L}_+)^{L+L_0}
\kket{QR,L;T} \ . 
\end{equation}
Of course, in the above expression, the normalization constant is omitted. 
General scheme for the above treatment is very complicated. 
In the next section, we will try to find a possible general scheme from a 
different viewpoint.

\section{Discussion}

In the previous section, we developed a method for constructing a possible 
orthogonal set for the $su(3)$-algebra and its associating $su(1,1)$-algebra. 
If we are concerned only with the $su(3)$-algebra, it may be enough to 
consider the case $T_0=T$:
\begin{equation}\label{5-1}
\kket{QR,LL_0;TT}=({\hat L}_+)^{L+L_0}\kket{QR,L;T} \ . 
\end{equation}
In the case restricted to $\{\kket{QR,LL_0;TT}\}$, the expression 
$\kket{Q^0,Q;L^0,LL_0}$ defined in the following form may be more 
understandable than $\kket{QR,LL_0;TT}$: 
\begin{equation}\label{5-2}
\kket{Q^0,Q;L^0,LL_0}=\kket{QR,LL_0;TT}\ . 
\end{equation}
As can be seen in the relations (\ref{3-3c}) and (\ref{3-3d}), we can 
use $L^0$ and $Q^0$ instead of $T$ and $R$.

Before showing some concrete results, we will discuss structure of the 
state $\kket{Q^0,Q;L^0,LL_0}$. 
As can be seen in the definition (\ref{2-9a}), $({\hat L}_{\pm,0})$ 
consists of two parts: 
One is expressed in terms of $({\hat a}_{\pm,0}^*,{\hat a}_{\pm,0})$ 
and the other $({\hat b}_{\pm,0}^*,{\hat b}_{\pm,0})$. 
Therefore, our problem is reduced to the coupling of two angular momenta. 
The quantum numbers $l_a$ and $l_b$ introduced in the relation 
(\ref{3-8a}) denote the magnitudes of the two angular momenta, respectively. 
Therefore, $L$ can run in the region 
\bsub\label{5-4}
\begin{equation}\label{5-4a}
L=l_a+l_b,\ l_a+l_b-1, \cdots , |l_a-l_b|+1, \ |l_a-l_b| \ . 
\qquad
(|l_a-l_b|\leq L\leq l_a+l_b)
\end{equation}
We can verify that under the condition (\ref{5-4a}), 
$\kket{nR,L;T}$ defined in the relation (\ref{4-7}) does not vanish. 
The condition (\ref{5-4a}) is rewritten as 
\begin{equation}\label{5-4b}
L=L^0,\ L^0-1, \cdots , |Q^0|+1, \ |Q^0| \ . 
\qquad
(|Q^0|\leq L\leq L^0)
\end{equation}
\esub

Next, let us search a possible form of the orthogonal set introduced 
in the relation (\ref{4-13a}): $\{\kket{kR,L;T}\}$. 
With the use of the form (\ref{4-7}), formally, $\kket{kR,L;T}$ can be 
expressed as 
\beqn\label{5-5}
\kket{kR,L;T}&=&
\sum_n D_{kn}\kket{nR,L;T} \nonumber\\
&=&\sum_n D_{kn}({\mib C}_1^*)^{2(T-3/2)-L-2n}({\mib C}_2^*)^n \ket{T,R}
\nonumber\\
&=&\kket{k;l_al_b,L} \ . 
\eeqn
Here, $D_{kn}$ denotes expansion coefficient with respect to linearly 
independent set $\{\kket{nR,L;T}\}$ and we used 
the relation (\ref{3-8}) for $(T,R)$ and $(l_a,l_b)$. 
We will try to obtain $\kket{k;l_al_b,L}$ in terms of an idea slightly 
different from the direct use of ${\mib C}_1^*$ and ${\mib C}_2^*$ 
defined in the relation (\ref{4-1}).

For the above-mentioned aim, first, we introduce the following two sets of 
operators: 
\bsub\label{5-6}
\beqn
& &{\mib A}_+={\hat a}_+^*{\hat a}_-^*-(1/2){\hat a}_0^{*2} \ , \qquad
{\mib A}_-={\hat a}_-{\hat a}_+-(1/2){\hat a}_0^{2} \ , \nonumber\\
& &{\mib A}_0=(1/2)({\hat a}_+^*{\hat a}_+ + {\hat a}_-^*{\hat a}_- 
+{\hat a}_0^*{\hat a}_0)+3/4 \ , 
\label{5-6a}\\
& &{\mib B}_+={\hat b}_+^*{\hat b}_-^*-(1/2){\hat b}_0^{*2} \ , \qquad
{\mib B}_-={\hat b}_-{\hat b}_+-(1/2){\hat b}_0^{2} \ , \nonumber\\
& &{\mib B}_0=(1/2)({\hat b}_+^*{\hat b}_+ + {\hat b}_-^*{\hat b}_- 
+{\hat b}_0^*{\hat b}_0)+3/4 \ . 
\label{5-6b}
\eeqn
The sets $({\mib A}_{\pm,0})$ and $({\mib B}_{\pm,0})$ form the 
$su(1,1)$-algebras. 
Of course, they are independent to each other. 
Further, we introduce the operator 
\begin{equation}\label{5-6c}
{\mib C}^*={\hat a}_0^*{\hat b}_-^*-{\hat a}_-^*{\hat b}_0^* \ . 
\end{equation}
\esub
Important relations which ${\mib A}_{\pm,0}$, ${\mib B}_{\pm,0}$ and 
${\mib C}^*$ obey are summarized as 
\bsub\label{5-7}
\beqn
& &
[\ {\hat L}_{\pm,0} \ , \ \hbox{\rm any\ of}\ ({\mib A}_{\pm,0})\ {\rm and}\ 
({\mib B}_{\pm,0})\ ]=0\ , 
\label{5-7a}\\
& &[\ {\hat L}_- \ , \ {\mib C}^*\ ]=0\ , \qquad 
[\ {\hat L}_0 \ , \ {\mib C}^*\ ]=-{\mib C}^* \ . 
\label{5-7b}
\eeqn
\esub

With the aid of the above relations, we will construct 
the state $\kket{k;l_al_b,L}$. 
First, the following relation should be noted: 
\beqn\label{5-8}
& &{\hat a}_+^*{\hat a}_+ + {\hat a}_-^*{\hat a}_- + {\hat a}_0^*{\hat a}_0 
=({\wtilde T}_0-3/2)+{\wtilde R}_0 \ , \nonumber\\
& &{\hat b}_+^*{\hat b}_+ + {\hat b}_-^*{\hat b}_- + {\hat b}_0^*{\hat b}_0 
=({\wtilde T}_0-3/2)-{\wtilde R}_0 \ .
\eeqn
Since the state (\ref{4-7}) is the eigenstate of ${\wtilde T}_0$ and 
${\wtilde R}_0$ with the eigenvalues $T$ and $R$, respectively, and we have 
the relation (\ref{3-8}), the state (\ref{5-5}), which we intend to search, 
should contain $l_a$ and $l_b$, respectively. 

Noting the above argument, we introduce the following state: 
\begin{equation}\label{5-9}
\kket{(e);m_am_b,l_al_b}=({\mib A}_+)^{m_a}({\mib B}_+)^{m_b}
\ket{(e);l_a-2m_a,l_b-2m_b} \ .
\end{equation}
Here, $\ket{(e);k_a,k_b}$ denotes 
\beqn\label{5-10}
& &\ket{(e);k_a,k_b}=({\hat a}_-^*)^{k_a}({\hat b}_-^*)^{k_b}\ket{0} \ , 
\nonumber\\
& &{\mib A}_-\ket{(e);k_a,k_b}={\mib B}_-\ket{(e);k_a,k_b}=0 \ . 
\eeqn
The state (\ref{5-9}) satisfies 
\beqn\label{5-11}
& &{\hat L}_-\kket{(e);m_am_b,l_al_b}=0 \ , \nonumber\\
& &{\hat L}_0\kket{(e);m_am_b,l_al_b}
=-(l_a+l_b-2(m_a+m_b))\kket{(e);m_am_b,l_al_b} \ . 
\eeqn
We can prove that even if $m_a'+m_b'=m_a+m_b$ for the case 
$(m_a'\neq m_a,\ m_b'\neq m_b)$, the states (\ref{5-9}) are orthogonal. 
Also, we note that the state $\kket{(e);m_am_b,l_al_b}$ obeys the 
relations (\ref{4-9b})$\sim$(\ref{4-11b}). 
Clearly, the eigenvalue for ${\hat L}_0$ is given as 
\begin{equation}\label{5-12}
L=l_a+l_b-2(m_a+m_b)=L^0-2m \ . \qquad (m=m_a+m_b)
\end{equation}
However, we have to point out the following two problems for 
$\kket{(e);m_am_b,l_al_b}$: 
(1) The state $\kket{(e);m_am_b,l_al_b}$ does not satisfy the relation 
(\ref{4-9a}) and (2) the difference between $L^0$ and $L$ is restricted to 
even number. 
The problem (1) is reserved for later argument: The following state plays 
an important role: 
\begin{eqnarray}\label{5-13}
\ket{(o);k_a,k_b}&=&
{\mib C}^*({\hat a}_-^*)^{k_a-1}({\hat b}_-^*)^{k_b-1}\ket{0} \nonumber\\
&=&{\mib C}^*\ket{(e);k_a-1,k_b-1} \ . 
\end{eqnarray}
The state $\ket{(o);k_a,k_b}$ obeys 
\begin{equation}\label{5-14}
{\mib A}_-\ket{(o);k_a,k_b}={\mib B}_-\ket{(o);k_a,k_b}=0 \ . 
\end{equation}
Then, we define the state 
\begin{equation}\label{5-15}
\kket{(o);m_am_b,l_al_b}=({\mib A}_+)^{m_a}({\mib B}_+)^{m_b}
\ket{(o);l_a-2m_a,l_b-2m_b} \ . 
\end{equation}
The state (\ref{5-15}) satisfies 
\beqn\label{5-16}
& &{\hat L}_-\kket{(o);m_am_b,l_al_b}=0 \ , \nonumber\\
& &{\hat L}_0\kket{(o);m_am_b,l_al_b}=-(l_a+l_b-2(m_a+m_b)-1)
\kket{(o);m_am_b,l_al_b} \ .
\eeqn
The state (\ref{5-15}) is also the eigenstate of ${\wtilde T}_0$ and 
${\wtilde R}_0$ with the eigenvalues $T=(1/2)(l_a+l_b)+3/2$ and 
$R=(1/2)(l_a-l_b)$, respectively. 
Of course, in the same meaning as that of 
$\{\kket{(e);m_am_b,l_al_b}\}$, $\{\kket{(o);m_am_b,l_al_b}\}$ is 
orthogonal. 
The eigenvalue of ${\hat L}_0$ is given as  
\begin{equation}\label{5-17}
L=l_a+l_b-2(m_a+m_b)-1=L^0-(2m+1) \ . \qquad (m=m_a+m_b)
\end{equation}
The above argument tells us that the state (\ref{5-15}) has the same 
properties as those of the state (\ref{5-9}): 
The relations (\ref{4-9b})$\sim$(\ref{4-11b}) are satisfied. 
The difference between $L^0$ and $L$ is odd number. 
However, the relation (\ref{4-9a}) is not satisfied. 
In this way, we could solve the problem (2).

Let us return to the problem (1). 
In order to get the set which satisfies the relation (\ref{4-9a}), we 
introduce the operator ${\wtilde P}$ in a form analogous to the form 
(I$\cdot$A$\cdot$1): 
\beqn
& &{\wtilde P}=1-{\wtilde T}_+({\wtilde T}_-{\wtilde T}_+)^{-1}{\wtilde T}_- 
\ , 
\label{5-18-0}\\
& &{\wtilde T}_-{\wtilde T}_+={\wtilde T}_0({\wtilde T}_0+1)
-{\wtilde {\mib T}}^2 \ . 
\label{5-18}
\eeqn
Since ${\wtilde T}_-{\wtilde T}_+$ is positive-definite in the form 
(\ref{2-10}), we can define $({\wtilde T}_-{\wtilde T}_+)^{-1}$. 
The operator ${\wtilde P}$ has the properties 
\bsub\label{5-19}
\beqn
& &{\wtilde P}^*={\wtilde P} \ , \qquad
{\wtilde P}^2={\wtilde P} \ , 
\label{5-19a}\\
& &{\wtilde T}_-{\wtilde P}={\wtilde P}{\wtilde T}_+=0 \ , \qquad
[\ {\wtilde T}_0 \ , \ {\wtilde P}\ ]=0 \ , \nonumber\\
& &[\ {\wtilde R}_0 \ , \ {\wtilde P}\ ]=0 \ , \qquad 
[\ {\hat L}_{\pm,0} \ , \ {\wtilde P}\ ]=0 \ . 
\label{5-19b}
\eeqn
\esub
The relation (\ref{5-19a}) tells us that ${\wtilde P}$ is a projection 
operator. 
Next, we define the state $\rkket{(\pm);m_am_b,l_al_b}$ in the form 
\begin{equation}\label{5-20}
\rkket{(\pm);m_am_b,l_al_b}={\wtilde P}\kket{(\pm);m_am_b,l_al_b}
\end{equation}
Operation of the form (\ref{5-19b}) on the state (\ref{5-20}) gives us 
\bsub\label{5-21}
\beqn
& &{\wtilde T}_-\rkket{(\pm);m_am_b,l_al_b}=0 \ , 
\label{5-21a}\\
& &{\wtilde T}_0\rkket{(\pm);m_am_b,l_al_b}=T\rkket{(\pm);m_am_b,l_al_b} \ , \ 
\label{5-21b}
\eeqn
\esub
\vspace{-0.8cm}
\begin{equation}\label{5-22}
{\wtilde R}_0\rkket{(\pm);m_am_b,l_al_b}=R\rkket{(\pm);m_am_b,l_al_b} \ , \ 
\end{equation}
\vspace{-0.8cm}
\bsub\label{5-23}
\beqn
& &{\hat L}_-\rkket{(\pm);m_am_b,l_al_b}=0 \ , 
\label{5-23a}\\
& &{\hat L}_0\rkket{(\pm);m_am_b,l_al_b}=-L\rkket{(\pm);m_am_b,l_al_b} \ . 
\label{5-23b}
\eeqn
\esub
We can see that the state $\rkket{(\pm);m_am_b,l_al_b}$ satisfies the 
relations (\ref{4-9a})$\sim$(\ref{4-11b}) and the problem (2) is 
solved. 
However, a new problem happens. 
We know the set $\{\kket{(\pm);m_am_b,l_al_b}\}$ forms an orthogonal set, 
but $\{\rkket{(\pm);m_am_b,l_al_b}\}$ does not form an orthogonal set. 
A simple example shows this fact: $(m_a=1,\ m_b=0)$ and 
$(m_a=0,\ m_b=1)$. 
Both states are not orthogonal. 
Therefore, by an appropriate method, we have to orthogonalize the set 
$\{\rkket{(\pm);m_am_b,l_al_b}\}$. 

For the above-mentioned aim, we remember the diagonalization of ${\mib Q}_0$ 
defined in the relation (\ref{4-4}). 
The operator ${\mib Q}_0$ satisfies 
\begin{equation}\label{5-24}
[\ {\mib Q}_0 \ , \ {\wtilde P}\ ]=0 \ . 
\end{equation}
In the space spanned by the set $\{\rkket{(\pm);m_am_b,l_al_b}; 
l_a=T-3/2+R,\ l_b=T-3/2-R,\ m_a+m_b=T-3/2-1/2-(1\mp 1)/4 \}$, 
we set up the eigenvalue equation in a form similar to the relation 
(\ref{4-17}): 
\begin{equation}\label{5-25}
{\mib Q}_0\kket{QR,L;T}=Q(2T-2)(2T-3/2)\kket{QR,L;T}\ . 
\end{equation}
Of course, the eigenstate $\kket{QR,L;T}$ is expanded in terms of 
the above-given orthogonal states. 
The property (\ref{5-24}) gives us 
\begin{equation}\label{5-26}
{\mib Q}_0{\wtilde P}\kket{QR,L;T}
=Q(2T-2)(2T-3/2){\wtilde P}\kket{QR,L;T} \ . 
\end{equation}
Therefore, if $\kket{QR,L;T}$ is an eigenstate of ${\mib Q}_0$, then, 
${\wtilde P}\kket{QR,L;T}$ is also the eigenstate. 
From the above argument, we can conclude that the orthogonal set for our 
aim is obtained by picking up the non-vanishing states for operation 
of ${\wtilde P}$.

%%%%%%%%%%%%%%%%%%
%\section*{Acknowledgements} 
%
%One of the authors (M. Y.) acknowledges to Professor T. Marumori, 
%%co-author of this paper, 
%who guided him in studying the algebraic aspects 
%of many-body systems, for example, as is shown in the first of 
%Ref. \citen{6}. 

%%%%%%%%%%%%%%%%%%%%%%%%%%%%%%%%%%%%%%%%%%%%%%%%%%%%%%%%%%%%%%%%%%%

\appendix
\section{Relation to the formalism given in (I)}

In this Appendix, we discuss the relation of the formalism given in this paper 
to that in (I). 
In (I), the $su(3)$-generators are denoted by $({\hat I}_{\pm,0},{\hat M}_0, 
{\hat D}_{\pm}^*, {\hat D}_{\pm})$. 
In this paper, the notations $({\hat L}_{\pm,0}, {\hat Q}_0, {\hat C}_2^*, 
{\hat C}_2, {\hat C}_1^*, {\hat C}_1)$ are used. 
They are related to 
\bsub\label{a1}
\beqn
& &{\hat I}_+=(1/2\sqrt{6})(\sqrt{3}{\hat L}_+-\sqrt{2}{\hat C}_1^*) \ , \qquad
{\hat I}_-=(1/2\sqrt{6})(\sqrt{3}{\hat L}_- -\sqrt{2}{\hat C}_1) \ , 
\nonumber\\
& &{\hat I}_0=(1/4)({\hat L}_0+{\hat Q}_0) \ , 
\label{a1a}\\
& &{\hat M}_0=(1/4)(3{\hat L}_0-{\hat Q}_0) \ , 
\label{a1b}\\
& &{\hat D}_+^*=(1/\sqrt{6}){\hat C}_2^* \ , \qquad
{\hat D}_-^*=(1/2\sqrt{6})(\sqrt{3}{\hat L}_+ +\sqrt{2}{\hat C}_1^*) \ , 
\nonumber\\
& &{\hat D}_+=(1/\sqrt{6}){\hat C}_2 \ , \qquad
{\hat D}_-=(1/2\sqrt{6})(\sqrt{3}{\hat L}_- +\sqrt{2}{\hat C}_1) \ . 
\label{a1c}
\eeqn
\esub
The boson operators used in this paper $({\hat a}_{\pm,0},{\hat b}_{\pm,0})$ 
correspond to $({\hat a}_{\pm}, {\hat a}, {\hat b}_{\pm}, {\hat b})$ 
in (I) as follows:
\beqn\label{a2}
& &{\hat a}_+ \longrightarrow {\hat a}_+ \ , \qquad 
{\hat a}_- \longrightarrow {\hat b} \ , \qquad 
{\hat a}_0 \longrightarrow {\hat a}_- \ , \nonumber\\ 
& &{\hat b}_+ \longrightarrow -{\hat a} \ , \qquad 
{\hat b}_- \longrightarrow {\hat b}_- \ , \qquad 
{\hat b}_0 \longrightarrow -{\hat b}_+ \ .
\eeqn
The $su(1,1)$-generators in (III) correspond to 
\beqn
({\wtilde T}_{\pm}, {\wtilde T}_0)_{III} &\longrightarrow& 
(-{\wtilde T}_{\pm}, {\wtilde T}_0)_I \ , 
\label{a3}\\
({\wtilde R}_0)_{III} &\longrightarrow& 
({\wtilde R}_0)_I \ . 
\label{a3-2}
\eeqn
Of course, the correspondence (\ref{a3}) is obtained under the relation 
(\ref{a2}). 
Inversely, we have 
\begin{equation}\label{a4}
{\hat L}_+=\sqrt{2}({\hat I}_+ + {\hat D}_-^*) \ , \qquad
{\hat L}_-=\sqrt{2}({\hat I}_- + {\hat D}_-) \ , \qquad 
{\hat L}_0={\hat I}_0+{\hat M}_0 \ . 
\end{equation}

In (I), we showed three forms for the eigenstate of 
${\wtilde {\mib T}}^2$, ${\wtilde T}_0$, ${\wtilde R}_0$, ${\hat {\mib I}}^2$, 
${\hat I}_0$ and ${\hat M}_0$ with the eigenvalues $T(T-1)$, $T_0$, 
$R=2I^0-(T-3/2)$, $I(I+1)$, $I_0$ and $M_0=3I^1-2(T-3/2)+I^0$, respectively. 
They are expressed in the form 
$\ket{I^1I^0,II_0;TT^0}$, $\dket{I^1I^0,II_0;TT^0}$ and 
$\rdket{I^1I^0,II_0;TT^0}$, respectively, which can be seen in the relations 
(I$\cdot$4$\cdot$10), (I$\cdot$5$\cdot$9) and (I$\cdot$A$\cdot$11), 
respectively. 
We denote the case $I_0=-I$ and $T^0=T$ as 
$\kket{I^1I^0,I;T}$. 
The relation (\ref{a4}) tells us 
\beqn
& &{\hat L}_0\kket{I^1I^0,I;T}=-L\kket{I^1I^0,I;T} \ , \quad (L\geq 0)
\label{a5}\\
& &-L=-I_0+M_0=-I+3I^1-2(T-3/2)+I^0 \ . 
\label{a6}
\eeqn
The relation (\ref{a6}) is rewritten as 
\begin{equation}\label{a7}
I=3I^1-2(T-3/2)+I^0+L \ . 
\end{equation}
Then, hereafter, we use the notation 
\begin{equation}\label{a8}
\kket{I^1I^0,I;T}=\rkket{I^1I^0,L;T}
\end{equation}
Of course, $\rkket{I^1I^0,L;T}$ satisfies 
\beqn
& &{\hat L}_-\rkket{I^1I^0,L;T}\neq 0 \ , 
\label{a9}\\
& &{\hat L}_0\rkket{I^1I^0,L;T}=-L\rkket{I^1I^0,L;T} \ . 
\label{a10}
\eeqn
The relations (\ref{a9}) and (\ref{a10}) tell us that the state 
$\rkket{I^1I^0,L;T}$ is the eigenstate of ${\hat L}_0$ with the eigenvalue 
$-L$, but not the eigenstate of ${\hat {\mib L}}^2$.

Our interest is to find the eigenstate of ${\hat {\mib L}}^2$. 
For this purpose, we introduce the following operator: 
\beqn\label{a11}
& &{\hat P}
=1-{\hat L}_+({\hat L}_-{\hat L}_+ + \varepsilon)^{-1}{\hat L}_- \ , 
\qquad 
{\hat L}_-{\hat L}_+={\hat {\mib L}}^2-{\hat L}_0({\hat L}_0+1) \ , \nonumber\\
& &\varepsilon\ : \ \hbox{\rm infinitesimal\ parameter}. 
\eeqn
The operator (\ref{a11}) and its properties are found in the relation 
(I$\cdot$A$\cdot$1), etc. 
Following the properties of ${\hat P}$, we obtain 
\bsub\label{a12}
\beqn
& &{\hat L}_-{\hat P}\rkket{I^1I^0,L;T}=0 \ , \nonumber\\
& &{\hat L}_0{\hat P}\rkket{I^1I^0,L;T}=-L{\hat P}\rkket{I^1I^0,L;T} \ , 
\label{a12a}\\
& &{\wtilde T}_-{\hat P}\rkket{I^1I^0,L;T}=0 \ , \nonumber\\
& &{\wtilde T}_0{\hat P}\rkket{I^1I^0,L;T}=T{\hat P}\rkket{I^1I^0,L;T} \ , 
\label{a12b}\\
& &{\wtilde R}_0{\hat P}\rkket{I^1I^0,L;T}=R{\hat P}\rkket{I^1I^0,L;T} \ . 
\quad (R=2I^0-(T-3/2))
\label{a12c}
\eeqn
\esub
Here, it should be noted that $I^1$ is a quantum number, which specifies 
the eigenvalue of ${\hat M}_0$, for the state $\rkket{I^1I^0,L;T}$. 
However, $I^1$ is not a quantum number for the state 
${\hat P}\rkket{I^1I^0,L;T}$. 
Therefore, under an appropriate orthogonalization, we can obtain the 
orthogonal set $\{|\!|kI^0,L;T)\!)\}$. 
Thus, we obtain 
\begin{equation}\label{a13}
|\!|kI^0,LL_0;TT_0)\!)=({\wtilde T}_+)^{T_0-T}({\hat L}_+)^{L+L_0}
|\!|kI^0,L;T)\!)\ . 
\end{equation}
The above is the relation of the formalism (III) to that of (I). 
This may be a possible translation of the original Elliott's idea in the 
present language.

\end{document}